\documentclass[a4paper]{jpconf}
\usepackage{graphicx}
\usepackage{subfigure,epsfig,amsmath,amsfonts,amssymb,xcolor,slashed,cite}
\usepackage[utf8]{inputenc}
\begin{document}
\title{Exotic $\Omega_c^0$ baryons from meson-baryon scattering}

\author{Gl\`oria~Monta\~na$^{1}$, \`Angels~Ramos$^{1}$ and Albert~Feijoo$^{2}$}

\address{$^{1}$ Departament de F\'{\i}sica Qu\`antica i Astrof\'{\i}sica and Institut de Ci\`encies del Cosmos (ICCUB), Universitat de Barcelona,
Mart\'{\i} i Franqu\`es 1, 08028 Barcelona, Spain}
\address{$^{2}$ Nuclear Physics Institute, 25068 \v Re\v z, Czech Republic}


\begin{abstract}
A meson-baryon interaction in the charm $+1$, strangeness $-2$ and isospin 0 sector is built from a t-channel vector meson exchange model employing effective Lagrangians. The implementation of coupled-channel unitarization in the s-wave scattering amplitudes gives rise to two structures that have similar masses and widths to those of the $\Omega_c(3050)^0$ and $\Omega_c(3090)^0$ states recently observed by the LHCb collaboration. A meson-baryon molecular interpretation of these resonances would assign their spin-parity to be $J^P=1/2^-$.
\end{abstract}

\section{Introduction}
The recent observation by the LHCb collaboration of five narrow $\Omega_c^0$ excited resonances \cite{Aaij:2017nav}
has triggered a lot of activity in the field of baryon spectroscopy aiming at understanding their inner structure and possibly establishing their unknown values of spin-parity. Quark models giving a $css$ quark content picture have been revisited \cite{Karliner:2017kfm,Wang:2017vnc,Wang:2017zjw,Chen:2017gnu,Chen:2017sci,Agaev:2017jyt,Cheng:2017ove,Wang:2017hej}
and pentaquark interpretations \cite{Huang:2017dwn,An:2017lwg,Kim:2017jpx} have also been investigated.

An alternative scenario is provided by models that can describe some of these resonances as quasi-bound states of an interacting meson-baryon pair \cite{Hofmann:2005sw,JimenezTejero:2009vq,Romanets:2012hm}, an approach that we have recently re-examined \cite{Montana:2017kjw} in view of the new experimental data. 
Similarly to the $P_c(4380)$ and $P_c(4450)$ excited nucleon resonances, for which a  pentaquark structure having a $c\bar{c}$ pair in its composition is more natural rather than being an extremely high energy excitation of the $3q$ system, it is also plausible to expect that some excitations in the $C=1$, $S=-2$ sector can be obtained by adding a $u\bar{u}$ pair to the natural $ssc$ content of the $\Omega_c^0$.  The hadronization of the five quarks could then lead to bound states, generated by the meson-baryon interaction in coupled channels. This possibility is supported by the fact that the $\bar K\Xi_c$ and $\bar K\Xi_c^\prime$ thresholds are in the energy range of interest, and that the excited $\Omega_c^0$ baryons under study have been observed from the invariant mass of the spectrum of $K^-\Xi^+_c$ pairs.

\section{Formalism}
The unitarized scattering amplitude $T_{ij}$ exhibiting dynamically generated resonances is obtained from solving the on-shell Bethe-Salpeter equation in coupled channels, 
\begin{equation}\label{eq:BSeq}
T_{ij}=V_{ij}+V_{il}G_{l}T_{lj},
\end{equation}
which implements the resummation of loop diagrams to infinite order.

The loop function built with the meson and baryon propagators,
\begin{equation}\label{eq.Gmatrix}
G_{l}=i\int \frac{d^4q}{(2\pi)^4}\frac{2M_l}{(P-q)^2-M_l^2+i\epsilon}\frac{1}{q^2-m_l^2+i\epsilon},
\end{equation}
is formally divergent and is regularized by means of the \textit{dimensional regularization} approach, which introduces the dependence on a subtraction constant $a_l(\mu)$ for each intermediate channel $l$ at a given regularization scale $\mu$ (see Eq.~(18) in \cite{Montana:2017kjw}).

The s-wave tree-level amplitude used as the kernel $V_{ij}$ is obtained from the t-channel vector meson exchange \cite{Hofmann:2005sw} that in the $t\ll m_V$ limit reduces to a contact Weinberg-Tomozawa (WT) term:
\begin{equation}\label{eq:Vij}
 V_{ij}(\sqrt{s})=-C_{ij}\frac{1}{4f^2}\left(2\sqrt{s}-M_i-M_j\right) N_i N_j
\end{equation}
where $M_i$, $M_j$ and $E_i$, $E_j$ are the masses and the energies of the baryons and the normalization factors are ${N=\sqrt{(E+M)/2M}}$.

The coefficients $C_{ij}$ are obtained from the evaluation of the t-channel meson-baryon interaction diagram introducing the effective Lagrangians of the hidden gauge formalism \cite{Hofmann:2005sw}:
\begin{equation}\label{eq:vertexVPP}
\mathcal{L}_{VPP}=ig\langle\left[\partial_\mu\phi, \phi\right] V^\mu\rangle,
\end{equation}
\begin{equation}\label{eq:vertexBBV}
\mathcal{L}_{VBB}=\frac{g}{2}\sum_{i,j,k,l=1}^4\bar{B}_{ijk}\gamma^\mu\left(V_{\mu,l}^{k}B^{ijl}+2V_{\mu,l}^{j}B^{ilk}\right),
\end{equation}
to describe the vertices coupling the vector meson to pseudoscalars (VPP) and baryons (VBB), respectively, in the pseudoscalar meson-baryon scattering, and assuming $SU(4)$ symmetry.

The interaction of vector mesons with baryons is built in a similar way and involves a three-vector $VVV$ vertex, which is obtained from the effective Lagrangian:
\begin{equation}\label{eq:vertexVVV}
\mathcal{L}_{VVV}=ig\langle {\left[V^\mu,\partial_\nu V_\mu\right] V^\nu}\rangle .
\end{equation}
The resulting interaction is that in Eq.~(\ref{eq:Vij}), including the product of polarization vectors, $\vec{\epsilon}_i\cdot\vec{\epsilon}_j$.

Note that, while $SU(4)$ symmetry is encoded in the Lagrangians, the interaction potential is not $SU(4)$ symmetric due to the use of physical masses for the mesons and baryons involved, as well as to a factor $\kappa_c=1/4$ accounting for the higher mass of the charmed mesons exchanged in some of the non-diagonal transitions.  Actually, the transitions mediated by the exchange of light vector mesons do not make explicit use of $SU(4)$ symmetry. This is for instance the case of the dominant diagonal transitions, which are effectively projected into their $SU(3)$ content.

The available pseudoscalar meson-baryon channels in the $(I,S,C)=(0,-2,1)$ sector are $\bar{K}\Xi_c (2964)$, $\bar{K}\Xi'_c (3070)$, $D\Xi (3189)$, $\eta \Omega_c (3246)$, $\eta' \Omega_c (3656)$, $\bar{D}_s \Omega_{cc} (5528)$, and $\eta_c \Omega_c (5678)$, where the values in parentheses indicate their corresponding threshold. The doubly charmed $\bar{D}_s \Omega_{cc}$ and $\eta_c \Omega_c $ channels will be neglected, as their energy is much larger than that of the other channels. 
The matrix of $C_{ij}$ coefficients for the resulting 5-channel interaction is given in Table~\ref{tab:coeff}.

\begin{table}[h]
\vspace{-0.15cm}
\caption{The $C_{ij}$ coefficients for the $(I,S,C)=(0,-2,1)$ sector of the  pseudscalar meson-baryon interaction.}
\lineup
\begin{center}
\begin{tabular}{l c c c c c}
\br
&{${\bar K}\Xi_c$}  & {${\bar K}\Xi_c^\prime$}  & { $D\Xi$}  & { $\eta\Omega_c^0$} &{$\eta^\prime\Omega_c^0$}  \\
\mr
{${\bar K}\Xi_c$}         & $1$ & $0$ & $\sqrt{3/2}~\kappa_c$  & $\m\m 0$ & $\m\m 0$     \\
{${\bar K}\Xi_c^\prime$}   &     & $1$ & $\sqrt{1/2}~\kappa_c$ & $-\sqrt{6}$ & $\m\m 0$   \\
{ $D\Xi$} &     &     &    $\m 2$        &  $-\sqrt{1/3}~\kappa_c$  & $-\sqrt{2/3}~\kappa_c$ \\
{ $\eta\Omega_c^0$} &     &     &     &  $\m\m 0$  &  $\m\m 0$\\
{ $\eta^\prime\Omega_c^0$}  &     &     &    &     &  $\m\m 0$   \\             
\br
\end{tabular}
\end{center}
\label{tab:coeff}
\vspace{-0.15cm}
\end{table}

In the vector meson-baryon case, the allowed states are $D^*\Xi (3326)$, $\bar{K}^*\Xi_c (3363)$, $\bar{K}^*\Xi'_c (3470)$, $\omega \Omega_c (3480)$, $\phi \Omega_c (3717)$, $\bar{D}_s^* \Omega_{cc} (5662)$ and $J/\psi \Omega_c (5794)$, where, again, we will neglect the doubly charmed states. The coefficients $C_{ij}$ can be straightforwardly obtained from those for the $PB$ interaction in Table~\ref{tab:coeff}, by considering the correspondences: $\pi \rightarrow \rho, K\rightarrow K^\ast, \bar{K}\rightarrow\bar{K}^\ast,$ $  D\rightarrow D^\ast, \bar{D}\rightarrow\bar{D}^\ast, {1/\sqrt{3}}\,\eta+\sqrt{2/3}\,\eta^\prime\rightarrow\omega$ and $-\sqrt{2/3}\,\eta+{1/\sqrt{3}}\,\eta^\prime\rightarrow\phi$.

A resonance generated dynamically from the coupled channel meson-baryon interaction appears as a pole of the scattering amplitude $T$ in the so-called {\it second Riemann sheet} of the complex energy plane. The coupling constants $g_i$ of the resonance to the various channels are obtained from the residues at the pole position $z_p$ while the compositeness, i.e., the amount of $i^{\rm th}$-channel meson-baryon component, is given by the real part of $-g_i^2(\partial G/\partial\sqrt{s})|_{z_p}$.

\section{Results and Discussion}
The Bethe-Salpeter equation in coupled channels of Eq.~(\ref{eq:BSeq}) has been solved using subtraction constant values, $a_l(\mu=1\rm~GeV)$, obtained by imposing the loop function to coincide, at the corresponding thresholds, with the loop function regularized with a cut-off $\Lambda=800\rm~MeV$. The corresponding pseudoscalar meson-baryon scattering amplitude shows two poles,
\begin{eqnarray}
M_1 =  {\rm Re}z_1= 3051.6~\!{\rm MeV},&\!\!\Gamma_1 = -2 {\rm Im}z_1= 0.45~\!{\rm MeV} \nonumber \\
M_2 =  {\rm Re}z_2 = 3103.3~\!{\rm MeV},&\!\!\!\!\!\Gamma_2 = -2 {\rm Im}z_1= 17~\!{\rm MeV\!,}
\label{eq:reson_800}
\end{eqnarray}
corresponding to resonances with spin-parity $J^P=1/2^-$. Their energies are very similar to the second and fourth $\Omega_c^0$ states discovered by LHCb \cite{Aaij:2017nav},  with properties:
\begin{eqnarray}
\Omega_c(3050)^0:~~& M=3050.2\pm0.1\pm0.1^{+0.3}_{-0.5}~{\rm MeV}, \nonumber \\
                                 &\Gamma=0.8\pm0.2\pm0.1~{\rm MeV},\nonumber \\
\Omega_c(3090)^0:~~& M=3090.2\pm0.3\pm0.5^{+0.3}_{-0.5}~{\rm MeV}, \nonumber \\
                                 &\Gamma=8.7\pm1.0\pm0.8~{\rm MeV}.
\label{eq:exp}
 \end{eqnarray}
 
Even if the mass of our heavier state is larger by 10 MeV and its width is about twice the experimental one, our results clearly show the ability of the meson-baryon dynamical models to generate states in the energy range of interest. In an attempt to explore the possibilities of our model, we let the values of the five subtraction constants vary freely within a reasonably constrained range and look for a combination that reproduces the characteristics of the two observed states, $\Omega_c(3050)^0$
and $\Omega_c(3090)^0$, within $2\sigma$ of the experimental errors.
For a representative set of $a_l(\mu=1\rm~GeV)$ with equivalent cut-off values in the range $[320-950]\;\rm MeV$ \cite{Montana:2017kjw}, the new properties of the poles are shown in Table~\ref{tab:pseudo}.
We note that the strongest change corresponds to $a_{\bar{K}\Xi_c}$, needed to decrease the width of the $\Omega_c(3090)^0$. Its equivalent cut-off value of 320 MeV is on the low side of the usually employed values but it is still naturally sized.

\begin{table}[hbt!]
\vspace{-0.2cm}
\caption{The $\Omega^0_c$ states generated employing vector-type WT zero-range interactions between a pseudoscalar meson and a ground state baryon, within a coupled-channel approach.}
\begin{center}
\lineup
\begin{tabular}{lcccc}
\br
\multicolumn{5}{c}{ {\bf $0^- \otimes 1/2^+$} interaction in the {\bf$(I,S,C)=(0,-2,1)$} sector } \\
\mr
&    \multicolumn{2}{c}{}    &   \multicolumn{2}{c}{} \\
  [-4.5mm]
$M\;\rm[MeV]$             &    \multicolumn{2}{c}{$\qquad 3050.3$}    &    \multicolumn{2}{c}{$\qquad 3090.8$}   \\
$\Gamma\;\rm[MeV]$   &     \multicolumn{2}{c}{$\qquad 0.44$}        &    \multicolumn{2}{c}{$\qquad 12$}     \\ 
\mr
 &    \multicolumn{2}{c}{}    &   \multicolumn{2}{c}{} \\
  [-3mm]
                                &      $\qquad | g_i|$    & $-g_i^2 dG/dE$    &   $\qquad | g_i|$  & $-g_i^2 dG/dE$ \\
$\bar{K}\Xi_c (2964)$   &  $\qquad 0.11$  & $0.00+i\,0.00$     &    $\qquad 0.49$  & $-0.02+i\,0.01$  \\
$\bar{K}\Xi'_c (3070)$  &  $\qquad 1.80$  & $0.61+i\,0.01$     &    $\qquad 0.35$  & $\m 0.02-i\,0.02$  \\
$D\Xi (3189)$           &  $\qquad 1.36$  & $0.07-i\,0.01$     &    $\qquad 4.28$  & $\m 0.91-i\,0.01$ \\
$\eta \Omega_c (3246)$  &  $\qquad 1.63$  & $0.14+i\,0.00$     &    $\qquad 0.39$  & $\m 0.01+i\,0.01$ \\
$\eta' \Omega_c (3656)$ &  $\qquad 0.06$  & $0.00+i\,0.00$     &    $\qquad 0.28$  & $\m 0.00+i\,0.00$ \\
\br
\end{tabular}
\end{center}
\label{tab:pseudo}
\vspace{-0.2cm}
\end{table}

Table~\ref{tab:pseudo} also displays the couplings of each resonance to the various meson-baryon channels and the corresponding compositeness. The lowest energy state at 3050~MeV couples appreciably to the channels $\bar{K}\Xi'_c$, $D\Xi $ and  $\eta \Omega_c$, being the coupling to  $\bar{K}\Xi'_c$ the strongest and also the compositeness in this channel. The higher energy resonance at 3090~MeV, with a strong coupling to $D\Xi$ and a compositeness in this channel of 0.90, clearly qualifies as being a $D\Xi $ bound state.

\begin{figure}[h]
\vspace{-0.15cm}
\includegraphics[width=0.5\textwidth]{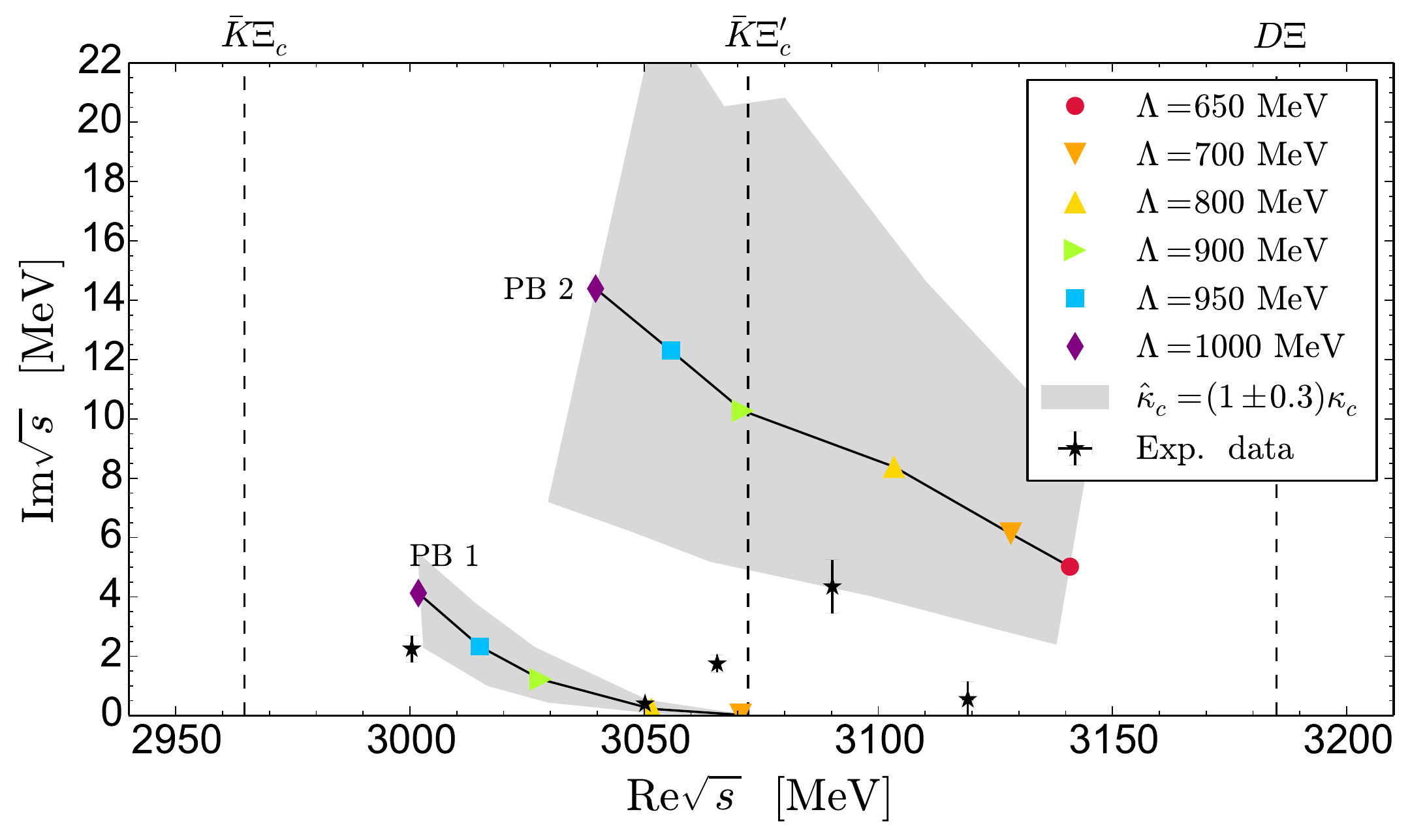}\hspace{0.5cm}
\begin{minipage}[b]{0.45\textwidth}\caption{Evolution of the pole resonance positions for various cut-off values. The grey area indicates the region of results covered when a variation of 30\% in the $SU(4)$ breaking is assumed in the transitions mediated by heavy-meson exchange.}\vspace{1cm}
\end{minipage}
\label{fig:cut-off}
\vspace{-0.15cm}
\end{figure}

Before proceeding further, we discuss the dependence of our results to the assumed value of the cut-off, as well as the influence of a certain amount of $SU(4)$ symmetry violation associated to the fact that the charm quark is substantially heavier than the light quarks. The solid lines in Figure~\ref{fig:cut-off} shows the evolution of the resonance poles as the value of the cut-off is varied between 650~MeV and 1000~MeV. As for implementing a violation of $SU(4)$ symmetry, we note that this is already partly done by the use of the physical meson and baryon masses in the interaction kernel. Moreover, the transitions mediated by light vector mesons will be left untouched, as only $SU(3)$ is effectively acting there. Therefore, we implement up to an additional $30\%$ of $SU(4)$ breaking only in the matrix elements that connect states via the t-channel exchange of a heavy vector meson, and this is effectively done by allowing the factor $\kappa_c$ to vary in the range $(0.7-1.3)\kappa_c$. The grey area in Figure~\ref{fig:cut-off} denotes the region in the complex plane where the resonances can be found varying the cut-off and the amount of $SU(4)$ breaking as described above. We see that the band of uncertainties includes the 3050~MeV and the 3090~MeV resonances measured at LHCb, a fact that gives strength to their interpretation as meson-baryon molecules.

A similar procedure has been followed to look for resonances in the vector meson-baryon scattering amplitude. Employing subtraction constants mapped onto a cut-off of $\Lambda=800$~MeV, we see a similar pattern as that found for the pseudoscalar-baryon case. A lower energy resonance, which mainly classifies as a $D^*\Xi$ bound state, is generated at 3231~MeV, while another resonance appearing at 3419~MeV is mainly a $\bar{K}^*\Xi'_c $ composite state with some admixture of  $\omega \Omega_c^0$ and $\phi \Omega_c$ components.  We find an additional pole in between these two, coupling strongly to $\bar{K}^*\Xi_c$ states. Each of these states corresponds to a degenerate $J^P=1/2^-,3/2^-$ doublet, yet the three of them are located at energy values well above the states found by the LHCb collaboration in a region where no narrow structures have been seen \cite{Aaij:2017nav}. We note, however, that the states found here from the vector-baryon interaction are artificially narrow as they do not couple to, and hence cannot decay into, the pseudoscalar-baryon states that lie at lower energy.

 Finally, we present predictions for the analogous bottom $\Omega_b^-$ resonances. These states are generated from meson-baryon interaction kernels obtained from the Lagrangians of Eqs.~(\ref{eq:vertexVPP}),(\ref{eq:vertexBBV}) and (\ref{eq:vertexVVV}), but replacing the charm mesons and baryons by their bottom counterparts. The matrices of coefficients are then the same as that in Table~\ref{tab:coeff}, but involving the following channels:
$\bar{K}\Xi_b (6289)$,
$\bar{K}\Xi'_b (6431)$,
$\eta \Omega_b (6594)$,
$\bar{B}\Xi (6598)$,
$\eta' \Omega_b (7004)$
for pseudoscalar-baryon states and
$\bar{B}^*\Xi (6643)$,
$\bar{K}^*\Xi_b (6687)$,
$\omega \Omega_b (6829)$,
$\bar{K}^*\Xi'_b (6829)$,
$\phi \Omega_b (7066)$
for vector-baryon ones. The coefficient analogous to $\kappa_c$ in certain non-diagonal transitions is $\kappa_b = 0.1$ to account for the much larger mass of the exchanged vector bottom mesons with respect to the light ones.

The results for the $\Omega_b^-$ resonances are very similar to those found in the charm sector. The interaction of pseudoscalar mesons with baryons generates two states with spin $J^P=1/2^-$, one at 6418~MeV, coupling strongly to ${\bar K}\Xi^\prime_c$ and $\eta\Omega_b$ states, and another at 6519~MeV, being essentially a $B\Xi$ bound state. The interaction of vector mesons with baryons generates $J=1/2^-,3/2^-$ spin degenerate $\Omega^-_b$ resonances at energies 6560~MeV, coupling strongly to ${\bar B}^*\Xi$, 6665~MeV, coupling strongly to $K^*\Xi_b$, and 6797~MeV, being a mixture of $\omega \Omega_b$, ${\bar K}^* \Xi'_b$ and $\Phi \Omega_b$.

\section{Conclusions}
We have studied the interaction of the low-lying pseudoscalar mesons with the ground-state baryons in the charm $+1$, strangeness $-2$ and isospin $0$ sector, employing a t-channel vector meson exchange model with effective Lagrangians. The resulting unitarized amplitude for the scattering of pseudoscalar mesons with baryons shows the presence of two resonances, having energies and widths very similar to some of the $\Omega_c^0$ states discovered recently at LHCb.

We have extended our study to the bottom sector and have found several $\Omega_b^-$ resonances in the energy region $6400-6800$~MeV that have a molecular meson-baryon structure.

The possible interpretation of some of the $\Omega_c$ states seen at CERN as being quasi-bound meson-baryon systems has been addressed in several other works \cite{Debastiani:2017ewu,Wang:2017smo,Chen:2017xat,Nieves:2017jjx}, as well as the prediction of analogous states in the bottom sector \cite{Liang:2017ejq}, finding results which are similar to the ones in our work \cite{Montana:2017kjw} and indicating that one cannot ignore the meson-baryon description when trying to understand the nature of these excited heavy baryons.

\ack
This work is supported by the Spanish MINECO under the contracts MDM-2014-0369 and FIS2017-87534-P and by the Generalitat de Catalunya under the doctoral grant $\rm{2018\;FI\_B\;00234}$.

\section*{References}

\bibliography{references}

\end{document}